# Dilepton production from p–p to Ca–Ca at the Bevalac[†]

H.S. Matis [a], S. Beedoe [b], M. Bougteb [c], J. Carroll [b], W. Christie [d], W. Gong [a], T. Hallman [d], L. Heilbronn [a], H. Huang [a], P.N. Kirk [e], G. Krebs [a], G. Igo [b], A. Letessier–Selvon [a], L. Madansky [d], F. Manso [c], J. Miller [a], C. Naudet [a], R.J. Porter [a], M. Prunet [c], G. Roche [a,c], L.S. Schroeder [a], P. Seidl [a], Z. F. Wang [e], R. Welsh [d], W.K. Wilson [a], and A. Yegneswaran [f]

(The DLS Collaboration)

[a] Lawrence Berkeley Laboratory, University of California, Berkeley, CA 94720 USA
[b] University of California at Los Angeles, Los Angeles, CA 90024, USA
[c] Université Blaise Pascal/IN2P3, F-63177 Aubrière Cedex, France
[d] Johns Hopkins University, Baltimore, MD 21218, USA
[e] Louisiana State University, Baton Rouge, LA 70803, USA
[f] CEBAF, Newport News, VA 23606, USA

The DLS collaboration has recently completed a high statistics study of dilepton production at the Bevalac. In particular, we have measured dielectrons ($e^+e^-$) from *p-p* and *p-d* collisions to understand the basic dilepton production mechanisms in the energy range from 1.05 - 4.9 GeV. These data can be used to determine the basic processes which contribute to nucleon-nucleon dilepton production such as hadronic bremsstrahlung, vector meson processes, and hadronic Dalitz decay. The data show that a simple elastic bremsstrahlung calculation is insufficient to explain the data. Theoretical models are compared with the data. A new high statistics study of *Ca-Ca* at 1.05 A·GeV has been made to study the collectivity of *A-A* collisions.

## 1. INTRODUCTION TO DILEPTON PRODUCTION

During the last several years, the Dilepton Spectrometer (DLS) collaboration[1,2,3,4,5] has studied the production of dileptons in both *p-A* collisions and *A-A* collisions. These publications demonstrated that there was a measurable dilepton cross section at Bevalac energies in the range of 1-2 A·GeV.

When nuclei collide, hadrons and real and virtual photons are produced. Hadrons strongly interact as they pass through the fireball until they escape. Therefore, we can say that these hadronic processes generally probe the late stages of the nuclear fireball. Both the virtual photon and the lepton pair from its decay interact only electromagnetically with the fireball, so that they essentially are unaffected after their production. Consequently, measurement of the virtual photon can provide a direct measurement of the interior of an *A-A* collision.

Gale and Kapusta[6] have first studied the process of pion annihilation in hot dense matter. In a static calculation, they have shown that dilepton production is sensitive to the interior density of a nucleus-nucleus collision. Furthermore, they demonstrated that the measurement of dileptons produced by $\pi$-$\pi$ annihilations should give information on the pion dispersion relationship in nuclear matter. This calculation has been extended to a dynamical model[7] by Xia *et al.* In recent papers, Gy. Wolf *et al.* have utilized[8] a BUU code while Winckelmann *et al.* have used a RQMD calculation to describe[9] dilepton data.

## 2. THE DLS EXPERIMENT- PHASE II

The data from the first phase of the DLS program clearly established a dilepton signal in systems which ranged from *p-Be* to *Nb-Nb*. The experimental errors were relatively large primarily due to the low statistics. Because of these large errors, these data sets were not selective enough to discriminate among competing theories. Therefore, the collaboration undertook substantial efforts to improve the amount and quality of the data. We also realized that it was necessary to measure the fundamental *N-N* cross sections in this energy region and to build a liquid $H_2$ and $D_2$ target. Many of these changes reflect improvements to the original DLS apparatus which was described in an earlier paper.[10]

**2.1 Operational and Experimental Improvements.**

Significant changes were made by Bevalac operations staff to insure we had adequate beam time in blocks of several weeks duration. The longer running periods significantly increased our operating efficiency and permitted more time to measure systematic effects. As the earlier data sets were usually taken during short three day periods, there was very little time to make careful systematic checks. In addition, during the last year of Bevalac operations, the duty factor was increased up to 90% from the previous 25%. This improvement provided many more interactions per unit time.

Beam wire chambers were used to monitor the beam several meters downstream of the target for each beam spill. Using this information, we were able quickly to alert the Bevalac operators to shifts in the beam position. Additionally, the data analysis program rejected all events from those spills that did not meet specific spatial characteristics. To improve the tracking performance of the spectrometer, extra planes were added to better measure tracks in the bending plane.

At the beginning of the high statistics phase of the experiment, it became apparent that there were significant efficiency effects which were dependent on the intensity of the beam. With that knowledge, several changes in the operating procedure of the experiment were made. The intensity of the beam for each energy, target, and projectile combination was carefully monitored. There was a conscious effort to keep the rates in the various detectors approximately the same for each energy, beam and target combination. Runs were made at different intensities to establish the rate dependence and this behavior was used to correct the data. The trigger electronics was modified to reduce the trigger rate dependence.

**2.2 Improvements to the DLS Acceptance Calculation.**

Significant changes have been made to the way that the DLS calculates the acceptance of the spectrometer. Due to increased computing power, we were able to increase the number of events generated by our Monte Carlo acceptance code. This resulted in a table with not only small bin sizes but also more statistical accuracy to calculate the acceptance. Furthermore, we were able to increase the phase space of the spectrometer by using an individual acceptance table for each of the four magnetic field configurations of the two DLS dipole magnets. The new table increases our coverage at low $p_t$ near 300 MeV/c. Upon application of the new acceptance table, the previous local maximum observed in our earlier data near M = 300 MeV/$c^2$ disappears. In this paper the new *Ca-Ca* data uses the new acceptance table and any comparison to theory must be done with the new DLS filter.

An assumption in creation of our acceptance table is that the virtual photon decays in its rest frame isotropicly in the angles $\phi$ and $\theta$ However, the DLS spectrometer has significant regions in *(y, $p_t$, m, $\phi$, $\theta$)* space where dileptons do not reach the detector so there is no acceptance. Since we cannot correct for these areas where there is no possible measurement, models with significant $f$ and $\hat{A}$ dependence need careful evaluation when they are compared to our data.

**3. PROTON-PROTON AND PROTON-DEUTERON PROGRAM**

Cross sections from the 4.9 GeV *p-p* and *p-d* data[11] have been analyzed and published. The p-p data plotted as a function of mass is shown in Figure 1. Prior to these measurements, it was widely claimed that the major contribution was due to the simple elastic bremsstrahlung. In this approximation, the yield from p-n collisions is much greater than from p-p collisions because the dipole contribution cancels in p-p. Consequently, dilepton production from p-d collisions should be significantly greater than p-p collisions. However our results,[12] some of which are shown in Figure 2, demonstrate that pair production from p-d to p-p is about two at 4.9 GeV. This implies that the p-p production mechanisms are roughly equal to the p-n production processes at this energy.

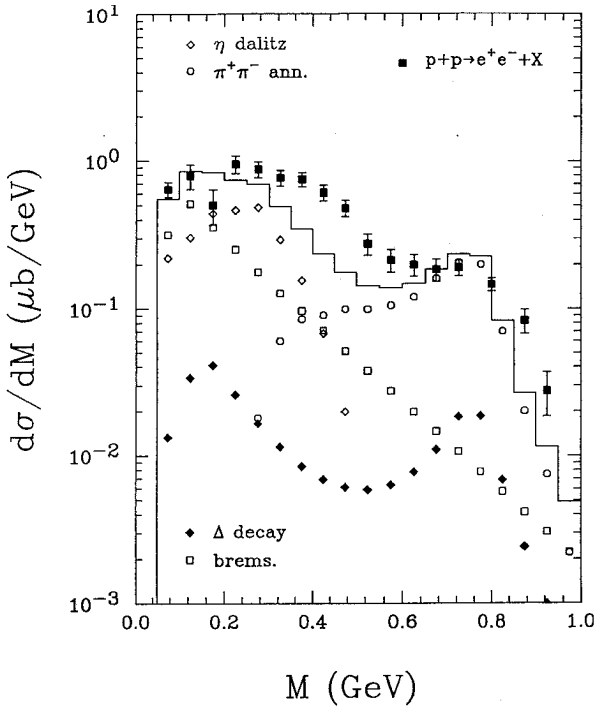

Figure 1. A comparison of DLS 4.9 GeV *p-p* data with calculation of Haglin and Gale. This figure is taken from Reference 13. The solid line is the sum of the various inelastic processes. Note that bremsstrahlung contributes at low mass.

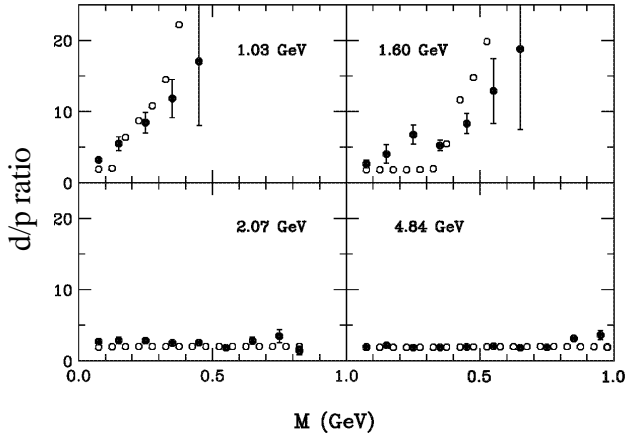

Figure 2. Comparison of the ratio of dilepton production for p-d and p-p collisions. The closed circles are DLS data points while the open circles are a calculation by Haglin. Pion annihilation is only included at the highest energy of the calculation.

Because of this discrepancy, Haglin and Gale made a more detailed calculation[13] of p-p and p-n elastic bremsstrahlung. Their results demonstrate that the p-p elastic cross section is actually larger than p-n in the DLS measurement region and both are several orders of magnitude smaller than the data. Superimposed on Figure 1 are the calculation of Haglin and Gale for inelastic bremsstrahlung, $\eta$ Dalitz decay and $\pi$-$\pi$ annihilation. Their calculation demonstrates that inelastic bremsstrahlung is a significant contributor to dilepton production at low mass. The addition of those processes approximately describes our data, except for a deficiency around 0.5 GeV/c$^2$. Processes such as $\omega$ Dalitz decay were not included in the calculation and may explain this difference.

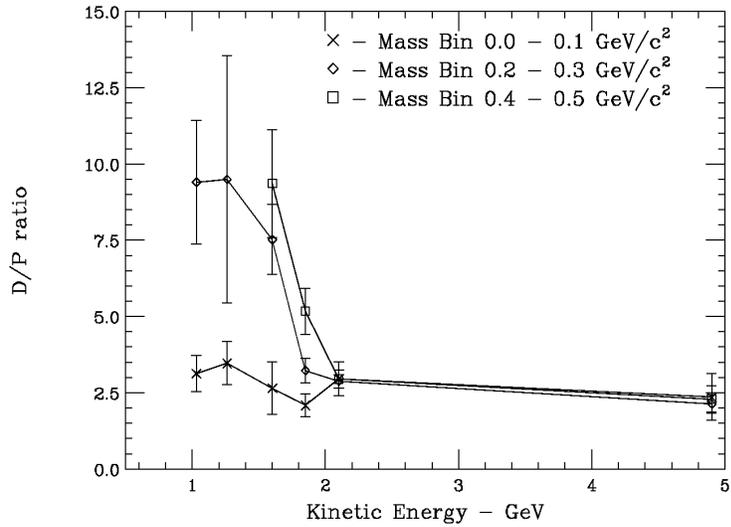

Figure 3. Excitation curve for different mass bins of the d/p ratio as a function of incident proton beam energy.

We measured the cross sections for *p-p* and *p-d* for several energies in the range between 1.0 to 4.9 GeV. Figure 2 shows the ratio of cross sections for four of the measured energies. Superimposed on our data is a calculation[14] by Haglin who uses the results of reference [12]. Another calculation by Schafer *et al.*,[15] using a fit to elastic nucleon-nucleon data, also reproduces the general behavior of the ratio data. The behavior of the ratio as a function of energy is shown in Figure 3. In the lowest mass bin the ratio is essentially constant as a function of the energy. The behavior of the ratio for higher mass bins can be explained by the presence of production thresholds, such as the $\eta$, in the 1-2 GeV energy region. At the lowest beam energy, *p-n* bremsstrahlung is much higher than *p-p* so that the *p-d* cross section is dominant.

## 4. NUCLEUS-NUCLEUS PROGRAM

A major goal of the DLS program is to establish whether any collective effects can be discerned in nucleus-nucleus collisions. Earlier work by our collaboration [3] indicated that there was a difference in the mass spectrum between *p-Be* and *Ca-Ca* data at 1.05 A·GeV. However, due to the low number of measured pairs for the earlier *Ca-Ca* system, 255 ± 35, a much more accurate measurement was needed.

The collaboration dedicated four weeks of data taking time to repeat the *Ca-Ca* measurement. At the present state of analysis, we have measured 2860 ± 128 *Ca-Ca* direct electron pairs. Due to the high multiplicity of these data, track pattern recognition needs to be carefully checked. In addition, several important calibrations and efficiency calculations need to be done and a proper normalization must be performed. Therefore, all results are preliminary. These data sets use the new DLS acceptance data and therefore any theoretical calculation will need to compare to the new DLS filter.

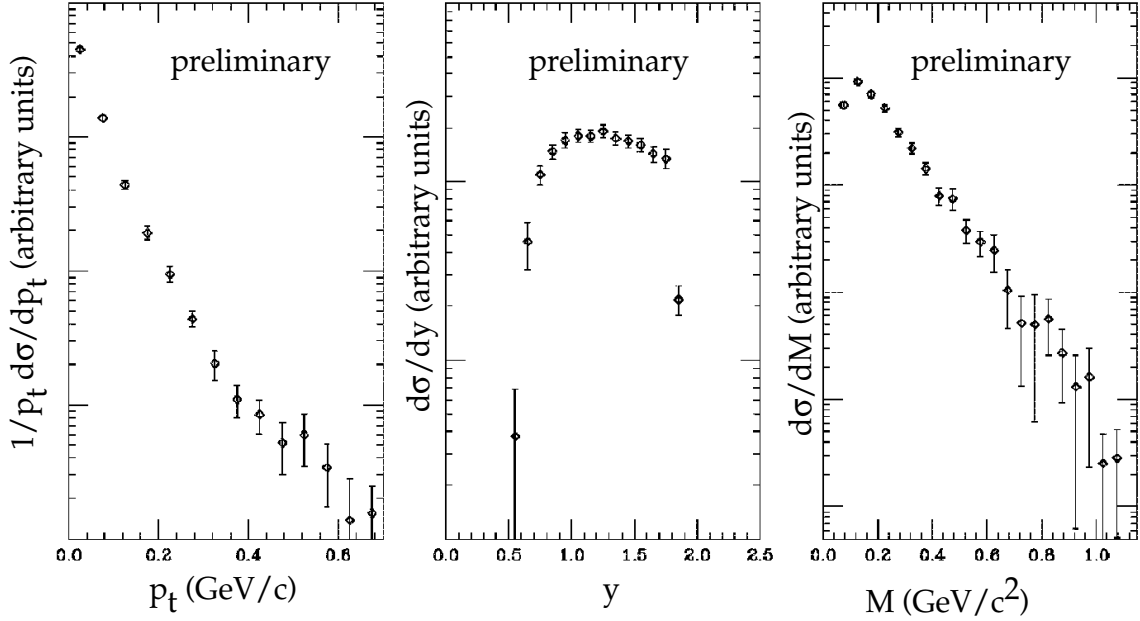

Figure 4. Preliminary results from the new Ca-Ca data. As with the other plots, the rapidity (y) spectrum includes only regions that we measure and therefore is affected by the DLS geometry. Its shape indicates the region where the DLS has measurement capability. The rapidity of the Ca-Ca center of mass is 0.69.

Figure 4 shows the $p_t$ (transverse momentum), y (rapidity), and M (mass) spectra for the Ca-Ca data with arbitrary normalization. Because of the much higher statistics, the spectra are much smoother than our previous measurement. It is important to note that the data are presented in three variables and all models need to be fit in those three variables simultaneously. Using the new acceptance table, there is no longer a local maximum near 300 MeV/$c^2$. The inverse slope of the Ca-Ca mass spectra (for 300 < M < 600 MeV/$c^2$) is 118 ± 8 MeV which agrees with our previous measurement of 125 ± 16 MeV for the same system. This slope is different from p-Be which we found to be 71 ± 18 MeV. This difference could be explained by enhanced effective energy due to Fermi motion and the increased contribution of $\pi$-$\pi$ annihilation in the hot dense matter.

## 5. FUTURE ANALYSIS

In addition to data presented in this paper, measurements have been made for several other systems. These results soon will be available. To determine whether there is any collective effect in Ca-Ca collisions, we measured both the d-Ca and $\alpha$-Ca system. The comparison of those small nuclear systems with the much larger Ca nucleus can determine whether there is any collective behavior in the Ca-Ca system. We also have studied a lighter symmetric mass system C-C. All of those studies were taken at 1.05 A·GeV. A multiplicity detector surrounded the target during all of the A-A measurements. This detector will be used to compare dilepton production as a function of impact parameter.

## 6. SUMMARY

The DLS experiment has evolved from a survey experiment which established the existence of a dilepton signal at Bevalac energies to a program which can provide the data needed to understand the production mechanisms. This study has benefited from a strong interaction with the nuclear theory community. While our analysis is not complete, a significant understanding of the simple *p-p* and *p-n* systems has been made. With the new high statistics A-A data, it should be possible to differentiate between several competing models.

[†] This work was supported in part by the Director, Office of Energy Research, Office of High Energy and Nuclear Physics, Nuclear Physics Division of the U.S. Department of Energy under contracts No. DE-AC03-76SF00098, No. DE-FG03-88ER40424, No. DE-FG02-88ER40413 and No. DE-FG05-88ER40445.

[1] G. Roche *et al.*, Phys. Rev. Lett., **61**,1069 (1988).

[2] C. Naudet *et al.*, Phys Rev. Lett., **62**, 2652 (1989).

[3] G. Roche *et al.*, Phys. Lett. **B226**, 228 (1989).

[4] S. Beedoe *et al.*, Phys. Rev. C **47**, 2840 (1993).

[5] H. S. Matis, contribution to the *Winter Workshop on Nuclear Dynamics VI*, LBL-28730 (1990).

[6] C. Gale and J. Kapusta, Phys. Rev. **C35**, 2107 (1987).

[7] L. H. Xia *et al.*, Nucl. Phys. **A485**, 721 (1988).

[8] Gy. Wolf e*t al.*, Nucl. Phys. **A552**, 549 (1993).

[9] L.A. Winckelmann, H. Stöcker, W. Greiner and H. Sorge, Phys Lett. **B298**, 22 (1993).

[10] A. Yegneswaran *et al.*, Nucl. Inst. and Methods **A290**, 61 (1990).

[11] H.Z. Huang *et al*., Phys. Rev. C **49**, 314 (1994).

[12] W. K. Wilson *et al.*, Phys. Lett. **B 316**, 245 (1993).

[13] K. Haglin and C. Gale, Phys. Rev. C **49**, 401 (1994).

[14] K. Haglin, in 9th. Annual LBL-GSI High Energy Heavy Ion Summer Study - October 1993, (1994).

[15] M. Schäfer, H. C. Dönges, A. Engel and U. Mosel, Nucl. Phys. **A** (in press) UGI-93-5 (1994).